\documentclass[twocolumn,epjc3]{svjour3}  
\smartqed  
\RequirePackage{graphicx}
\journalname{Eur. Phys. J. C}
\usepackage{cite}
\usepackage{amsmath}
\usepackage{amsfonts}
\usepackage{amssymb}
\usepackage{makeidx}
\usepackage{graphicx}
\usepackage{lmodern}
\usepackage{color}
\DeclareMathOperator{\Tr}{Tr}

\usepackage{widetext}
\begin{document}
\title{\textbf{{\color{blue} Neutrino masses in an $SU(4)\times U(1)$ -- electroweak model with a scalar decuplet.
}}}
%

\author{Nguyen Anh Ky\thanksref{e1,addr1}
        \and 
        Nguyen T. Hong Van\thanksref{e2,addr2, addr3}
        \and Dinh Nguyen Dinh\thanksref{e3,addr2}
        \and 
        Phi Quang Van\thanksref{e4,addr2}
}
%
\thankstext{e1}{e-mail: anhky AT iop.vast.ac.vn}
\thankstext{e2}{e-mail: nhvan AT iop.vast.ac.vn}
\thankstext{e3}{e-mail: dndinh AT iop.vast.ac.vn}
\thankstext{e4}{e-mail: pqvan AT iop.vast.ac.vn}

\institute{\small
\textit{Institute of research and development, Duy Tan
university,\\ 
Danang 550000, Vietnam.\\}
\label{addr1}
 \and
\small
\textit{International centre for physics and Center for theoretical physics\\
at Institute of physics},
\small
\textit{Vietnam academy of science and technology},\\ 
\small
\textit{10 Dao Tan, Ba Dinh, Hanoi, Viet Nam.\\}\label{addr2}
          \and
          \small
\textit{Institute for interdisciplinary research in science and education},\\ 
\small
\it{ICISE, Quy Nhon, Viet Nam.\\}
\label{addr3}
}

\date{Received: date / Accepted: date}
%
%
\maketitle
\begin{abstract}
A neutrino mass model is suggested within an $SU(4)\times U(1)$ -- electroweak theory. 
The smallness of neutrino masses can be guaranteed by a seesaw mechanism realized through   
Yukawa couplings to a scalar $SU(4)$-decuplet. In this scheme the light active neutrinos are 
accompanied by heavy neutrinos, which may have masses at different scales, including those 
within eV-MeV scales investigated quite intensively in both particle physics and astrophysics/cosmology. 
The flavour neutrinos are superpositions of light neutrinos and a small fraction of heavy 
neutrinos with the mixing to be determined by the model's parameters (Yukawa coupling 
coefficients or symmetry breaking scales). The distribution shape of the Yukawa couplings  
can be visualized via a model-independent distribution of the neutrino mass matrix 
elements derived by using the current experimental data. 
The absolute values of these Yukawa couplings are able to be determined if the symmetry breaking scales are known, and vice versa.
With reference to several current and near future experiments, detectable bounds of these 
heavy neutrinos at different mass scales are discussed and estimated.    
\end{abstract}
~

\maketitle
\section{\label{sec:level1}Introduction}

\noindent

Particle physics is experiencing a special period when different big experiments have
been carried out and announced remarkable results, especially, after the discovery of
a scalar boson (called the Brout-Englert-Higgs boson or, briefly, Higgs boson), which
is likely the last puzzle piece filling up the particle content of the standard model 
\cite{Aad:2012tfa,Chatrchyan:2012ufa} (for a review, see, for example, \cite{Ky:2015eka}).
Thus, the standard model (SM) \cite{quangyem} proves once again to be an
excellent model of elementary particles and their interactions as it can explain various
phenomena and many its predictions have been confirmed by the experiment. However, there 
are a number of problems remaining unsolved by the SM and showing that the latter could 
be just an effective low-energy appearance of a high-energy theory. Neutrino masses and 
mixing 
\cite{Fukuda:1998tw,Fukuda:1998ub,Fukuda:1998mi,Ahmad:2001an,Ahmad:2002jz,Ahmad:2002ka}
are one of such problems calling for a modification of the SM. This problem is important 
not only in particle physics but also in other fields of physics such as nuclear physics, 
astrophysics and cosmology \cite{Bilen,Carlo,Mohapatra:1998rq,Lesgou}. Many models of 
neutrino masses and mixing have been proposed but none of them has been recognized as the 
right model yet. It is why we continue to look for other possibilities leading to building 
different models beyond the SM. There are several methods for building an extended SM, 
but the first and, maybe, most-often used one is that of extending the SM gauge group 
$SU(3)_c\otimes SU(2)_L \otimes U(1)_Y$ to a larger gauge group. A simpler method is 
to enlarge only the electro-weak part $SU(2)_L \otimes U(1)_Y$
of the SM gauge group to, for example, $SU(3)_L \otimes U(1)_X$ (the 3-3-1 model)
\cite{Fritzsch:1976dq,Pisano:1991ee,Frampton:1992wt,Foot:1992rh,Singer:1980sw,Valle:1983dk,
Montero:1992jk} or $SU(4) \otimes U(1)_X$ (the 3-4-1 model)
\cite{Voloshin:1987qy,tuan341,Pleitez:1993vf,Foot:1994ym,Pisano:1994tf}. 
These models have attracted interest of a number of authors for over 20 years because of 
their relative simplicity. However, compared with the 3-3-1 model, the 3-4-1 model has been 
less investigated (one of the reasons might be the 3-4-1 model has a bigger gauge group, thus 
it is more complicated)
but the latter has a richer structure which may provide more chance to explain the beyond
SM phenomenology. The 3-4-1 model was first introduced by M. Voloshin in \cite{Voloshin:1987qy}
and re-considered later by other authors (see, for example,
\cite{tuan341,Pleitez:1993vf,Foot:1994ym,Pisano:1994tf}). Originally, this model is characterized
by fermions (leptons or quarks) in each family grouped in an $SU(4)$-quartet (or quartet for short), 
and its scalar sector composed often of quartets, an $SU(4)$-decuplet (decuplet) and, sometimes, 
also an $SU(4)$-sextet (or the corresponding anti-multiplets). 
Above, in particular, the term ``$SU(4)$-quartet" means a quartet $\mathbf{4}$ or anti-quartet $\mathbf{4^*}$. For an anomaly cancellation \cite{Dobrescu:2001ae,Nisperuza:2009xm} the model requires an equal number of  $\mathbf{4}$ and $\mathbf{4^*}$ in the fermion sector. One of the possible variants is to choose all the lepton families and one of the quark families, say, the third one, to transform as $\mathbf{4}$, while the remaining two quark families to transform as $\mathbf{4^*}$, provided that the number of either families or colors is 3 (see, for example, \cite{Nisperuza:2009xm} for an anomaly-free structure of fermion sector of an 3-4-1 model). Here considering  neutrino masses and mixing only, we temporarily put the quark sector aside.
\\

In comparison with the SM (and the 3-3-1 model) the 3-4-1 model has a bigger particle
content, including an extended scalar sector, providing more possibilities for solving
different problems, in particular, that of neutrino masses and mixing (the price is the
introduction of more parameters). Especially, an extended scalar sector may provide a 
richer structure of neutrino masses. However, the problem of neutrino masses and mixing, 
so far, has not been investigated very much within the 3-4-1 model, moreover, to our 
knowledge, such an investigation using a scalar decuplet (decuplet, for short) is still poor, in particular, 
a seesaw mechanism based on a decuplet has not yet been considered. The present paper is also motivated by noticing that in the 3-4-1 model the VEV configuration of a decuplet can provide a seesaw structure and the seesaw mechanism can 
be automatically applicable at the leading order by Yukawa coupling to only a single decuplet (with an appropriate VEV), 
unlike in most of other models, where the seesaw mechanism usually requires more scalar multiplets  involved.
 Another motivation to use a decuplet for generating neutrino masses is that the latter 
(as well as charged lepton masses) can not be generated directly at the leading order by using quartets which are fundamental representation multiplets of the gauge group $SU(4)$. 
\\

As said above, neutrinos are massive but, according to the current particle physics 
experimental data and cosmological observation constraints
\cite{Tanabashi:2018oca,Palanque-Delabrouille:2015pga,Loureiro:2018pdz,
Couchot:2017pvz,Mertens:2016ihw,Aghanim:2018eyx}, their masses are very tiny, 
just of the order of $10^{-1} eV$, even less. Thus, one must find a way to explain that. 
One of the most popular ways to generate neutrino small masses is based on the so-called 
see-saw mechanism (there is a vast literature on this matter but one can see, for example,  
\cite{GellMann:1980vs,Mohapatra:1979ia,Schechter:1980gr,Schechter:1981cv} for the type-I 
see-saw mechanism and \cite{Bilen,Carlo,Mohapatra:1998rq} for a review on further developments). 
This mechanism has been applied to the SM and many extended models, in particular, to our 
knowledge, it was applied for the first time to the 3-3-1 model with right-handed neutrinos 
by using a scalar $SU(3)_L$-sextet in \cite{Ky:2005yq,Dinh:2006ia}. The latter papers 
inspire the present work and a later work, showing that the seesaw mechanism can be applied 
to the 3-4-1 model with and without a decuplet. One of the feature of the seesaw mechanism 
is the presence of one or more right-handed neutrinos which are ``naturally" introduced in 
the 3-4-1 model as fundamental representation (quartet) partners of right-handed charged 
leptons (it is an advantage of this model as in most of other models, except a few ones 
like those based on the left-right symmetry 
\cite{Pati:1974yy, Mohapatra:1974hk,Mohapatra:1974gc,Mohapatra:1979ia,Mohapatra:1980yp}, 
the right-handed neutrinos are introduced ``artificially" by hand). Now let us first make 
a quick introduction to the 341 model.\\

The plan of this article is the following. In the next section a concise  introduction to 
the 3-4-1 model with a concentration on its lepton and scalar sectors is presented. Section 
III is devoted to using an $SU(4)$ decuplet scalar for generation of neurtino masses. Some 
comments and conclusions are made in the final section. 

\section{The 3-4-1 model in brief}

\noindent 

This extended standard model is based on the gauge group $SU(3)_c\otimes SU(4) \otimes U(1)_X$. 
The latter is attractive by several reasons such as 
in this model two lepton chiralities of each family are unified in a fundamental representation 
of the $SU(4)$ gauge group and this model, similar to the 331 model, can explain the number of 
fermion families to be three \cite{Pisano:1991ee,Frampton:1992wt,Dobrescu:2001ae,Nisperuza:2009xm}.  
Because the subject of the present paper is neutrino masses we will consider only the lepton- 
and the scalar sectors of the model and leave its gauge- and quark sectors for a future 
research. As in the case of the 3-3-1 model, the 3-4-1 model, depending on the particle content 
and their alignment, has several versions. Let us consider one of the possible versions.

\subsection{Lepton sector}

\noindent

Many neutrino mass models require the introduction of right-handed neutrinos (the number of
which depends on the model considered), here we work in a model with a right-handed neutrino
(RHN), say $N_R^\alpha$, introduced for each family $\alpha=e,\mu,\tau$. As usually, 
these RHN's
are sterile neutrinos being singlets under the electroweak gauge group $SU(2)_L \otimes U(1)_Y$.
One of the main features of the 3-4-1
model is all leptons in each (extended) family are grouped in an $SU(4)$ quartet. An alignment
of these quartets can be
\begin{equation}f^a_L =
\left[\begin{array}{c} \nu_L^\alpha\\[2mm] l_L^\alpha\\[2mm]
(N_R^\alpha)^c\\[2mm]
(l_R^\alpha)^c
\end{array}\right], ~~\alpha=e,\mu,\tau,
\label{lepton}
\end{equation}
where $\nu_L^\alpha$ and $N_R^\alpha$ are neutrino fields (left- and right handed, respectively),
$l_L^\alpha$ and $l_R^\alpha$ are charged lepton fields, and $\alpha$ is a family (flavour) index,
while $F^c$ denotes the charge conjugation of a field $F$. In this model, $N_R^\alpha$ are
sterile neutrinos by introduction and can be replaced by arbitrary sterile/exotic leptons to make other models. The
transformation of $f^\alpha_L$, being also an $SU(3)_c$-singlet and $U(1)_X$-neutral, under
$SU(3)_c\otimes SU(4) \otimes U(1)_X$ is summarized as follows
\begin{equation}
f^\alpha_L \sim (1,4,0).
\end{equation}

Another alignment of the lepton multiplet, 
\begin{equation}f'^\alpha_L = \left[\begin{array}{c}
\nu_L^\alpha\\[2mm]
l_L^\alpha\\[2mm]
(l_R^\alpha)^c\\[2mm]
(N_R^\alpha)^c
\end{array}\right] \sim (1,4,0),~~\alpha=e,\mu,\tau,
\label{lepton2}
\end{equation}
is obtained from the one in \eqref{lepton} by exchanging the positions 
of the third and the fourth components. Working with which alignment among \eqref{lepton} 
and \eqref{lepton2} is the question of convenience depending on the choice of a gauge symmetry 
breaking scheme. For example, if we want the 3-4-1 model to be broken to the 3-3-1 model with 
two neutrinos in a lepton $SU(3)_L$ triplet \cite{Singer:1980sw,Valle:1983dk,Montero:1992jk} 
or the minimal 3-3-1 model \cite{Pisano:1991ee,Frampton:1992wt,Foot:1992rh}, we choose the 
alignment \eqref{lepton} or the alignment \eqref{lepton2}, respectively. In this paper the 
alignment \eqref{lepton} is chosen. Other versions of the 3-4-1 model, in which the third 
and the fourth components of an $SU(4)$ quartet \eqref{lepton} or \eqref{lepton2} are 
occupied by other leptons such as exotic charged leptons and arbitrary sterile neutrinos, 
could be also considered.
\\

To generate neutrino masses we must introduce an appropriate scalar sector. It can have
different structures but below we will work with that containing an $SU(4)$ decuplet.

\subsection{Scalar sector}

\noindent

Let us consider a scalar sector of the 3-4-1 model with three quartets,

\begin{eqnarray}
\eta&=& \left[ \begin{array}{c}
\eta_1^0 \\[2mm] \eta_2^- \\[2mm] \eta_3^0\\[2mm]\eta_4^+
\end{array} \right]
\sim (1,4,0),~~
\rho= \left[ \begin{array}{c}
\rho_1^+ \\[2mm] \rho_2^0 \\[2mm] \rho_3^+\\[2mm]\rho_4^{++}
\end{array} \right]
\sim(1,4,1),\nonumber\\[4mm]
\chi&=& \left[ \begin{array}{c}
\chi_1^{-} \\[2mm] \chi_2^{--} \\[2mm] \chi_3^-\\[2mm]\chi_4^0
\end{array} \right]
\sim (1,4,-1),
\end{eqnarray}
and one decuplet,
\begin{equation}
{\Delta}\sim \left[ \begin{array}{cccc}
\Delta_{11}^0 & \Delta_{12}^- & \Delta_{13}^0 & \Delta_{14}^+\\[3.8mm]
\Delta_{12}^- & \Delta_{22}^{--} & \Delta_{23}^- & \Delta_{24}^0\\[3.8mm]
\Delta_{13}^0 & \Delta_{23}^- & \Delta_{33}^0 & \Delta_{34}^{++}\\[3.8mm]
\Delta_{14}^+ & \Delta_{24}^0 & \Delta_{34}^{++} & \Delta_{44}^+
\end{array} \right]
\sim (1,10,0).
\label{10tet}
\end{equation}
In \eqref{10tet} the normalisation coefficients which
can be found by using the kinetic term of ${\Delta}$ are skipted. Sometimes,
the scalar sector is extended with one more quartet similar to $\eta$, say, 
\begin{equation}
\xi= \left[ \begin{array}{c}
\xi_1^0 \\[2mm] \xi_2^- \\[2mm]
\xi_3^0\\[2mm]\xi_4^+
\end{array} \right]
\sim (1,4,0),
\end{equation}
in order to resolve a quark mass problem \cite{Rodriguez:2007jc} or/and with
a (self-conjugate) sextet if a neutrino magnetic moment is included in
consideration \cite{Voloshin:1987qy}. Adding the scalar $\xi$ could be also
motivated by the fact that $\eta$ has two neutral components which may need
two independent vacuum (VEV) structures \cite{Ky:2005yq,Dinh:2006ia}. The
scalar sector containing only quartets has been used in different investigations
without giving fermion masses at the Yukawa coupling tree levels. The decuplet
\cite{Voloshin:1987qy} is introduced to generate charged lepton masses (with
the presence of only the sextet some of the charged leptons remain massless)
but it seems, it has not been used much for the neutrino mass generation. We
will explore the latter in this paper following an idea close to that of
\cite{Ky:2005yq,Dinh:2006ia}.\\

 For further use the VEV's of the scalars are denoted as follows.
 \begin{eqnarray}
\langle \eta \rangle &=& \left[
u_1,0, u_3,0
\right]^T,~~\langle \xi \rangle = \left[
v_1,0,v_3,0
\right]^T,\nonumber\\[2mm]
\langle \rho \rangle &=& \left[0,\sigma_2,0,0
\right]^T, ~~ \langle \chi \rangle = \left[0, 0, 0,w_4
\right]^T,
\label{vev4tet}
 \end{eqnarray}
\begin{equation}
\langle {\Delta}\rangle = \left[ \begin{array}{cccc}
\delta_1 ~&~ 0  ~&~ \delta_2 ~&~ 0\\[.9mm]
0 ~&~ 0 ~&~ 0 ~&~ \delta_4\\[.9mm]
\delta_2 ~&~ 0 ~&~ \delta_3 ~&~ 0\\[.9mm]
0 ~&~ \delta_4 ~&~ 0 ~&~ 0
\end{array} \right].
\label{vev10tet}
\end{equation}
For the sake of completeness, a sextet can be also introduced,
\begin{equation}
{\cal S}\sim \left[ \begin{array}{cccc}0 & -S_{12}^- & -S_{13}^0 & -S_{14}^+\\[3.8mm]
S_{12}^- & 0 & -S_{23}^-& -S_{24}^0\\[3.8mm]
S_{13}^0 & S_{23}^- & 0& -S_{34}^+\\[3.8mm]
S_{14}^+ & S_{24}^0 & S_{34}^+& 0
\end{array} \right]
\sim (1,6,0),
\label{6tet}
\end{equation}
with the VEV
\begin{equation}
\langle {\cal S}\rangle = \left[ \begin{array}{cccc}
0 ~& 0 & -s_1 & ~0\\[1.2mm]
0 ~& 0 & ~0& -s_2\\[1.2mm]
s_1 ~& 0 & ~0 & ~0\\[1.2mm]
0 ~& ~s_2 & ~0 & ~0
\end{array} \right].
\label{vev6tet}
\end{equation}
Below we will see that to generate neutrino masses (and also masses of
other leptons) at the tree level no quartet and sextet but only decuplet 
is relevant.
\section{Decuplet and neutrino mass generation}

\noindent

A neutrino mass generation can be realized by coupling $\bar{f}_L(f_L)^c$ to scalars
transforming under appropriate representations of $SU(4)$. Since both $\bar{f}_L$
and $(f_L)^c$ transform as an anti-quartet $\mathbf{4^*}$, their product
$\bar{f}_L(f_L)^c$ transforms as ${\mathbf{4^*}}\otimes {\mathbf{4^*}}$, which in turns
can be decomposed as a direct sum of a anti-sextet (which could be self-conjugate) an
anti-decuplet: 
${\mathbf{4^*}}\otimes {\mathbf{4^*}}= \mathbf{6^*} \oplus \mathbf{10^*}$.
Therefore, a scalar coupled to $\bar{f}_L(f_L)^c$ must tranform as $\mathbf{6}$
or $\mathbf{10}$, thus, it can be a sextet \eqref{6tet} or a decuplet \eqref{10tet}.
Thus, the Yukawa coupling of $\bar{f}_L(f_L)^c$ to scalars has the general form 
(cf. \cite{Voloshin:1987qy})
\begin{equation}
-{\cal L}_{f\cal SD}=
Y_{\alpha\beta}^S\bar{f}^\alpha_L({f^\beta_L})^c {\cal S}
+ Y_{\alpha\beta}^\Delta\bar{f}^\alpha_L({f^\beta_L})^c {\Delta},
\label{YSD}
\end{equation}
where $Y_{\alpha\beta}$ are coupling coefficients with 
$\alpha$ and $\beta$ being family indices which in general may not 
coincide with those of the real charged-lepton mass states but 
it is easy to see that we can work in the basis labeled by the latter, 
$\alpha,\beta =e,\mu,\tau$, starting from \eqref{lepton}. 
The leptons may get masses when the scalars in \eqref{YSD} develop VEV's.
Since the coupling to the sextet in \eqref{YSD} cannot provide a right 
lepton mass term it is discarded from consideration here, 
while the coupling to the decuplet can give lepton-mass-like terms, namely, 
a charged-lepton mass term if $\delta_4\neq 0$, and, in some circumstance 
(see below), a neutrino mass term via a see-saw mechanism.  
The latter is very important because it can generate small neutrino masses, 
accompanied, though, by a large mass scale (of heavy hypothesized neutrinos). 
Thus, for a generation of lepton masses instead of \eqref{YSD}
we have 
\begin{equation}
-{\cal L}_{f\Delta}= Y_{\alpha\beta}^\Delta\bar{f}^\alpha_L({f^\beta_L})^c {\Delta}, 
~~ \alpha,\beta =e,\mu,\tau.
\label{YD}
\end{equation}
However, using only the decuplet as in \eqref{YD} to generate masses of both 
charged leptons and neutrinos may lead to a wrong correlation between these 
masses 
(as the charged lepton- and neutrino mass matrices, which in this case are 
proportional to the same Yukawa matrix, can be diagonalized by the same 
unitary matrix, the PMNS matrix is trivial). 
It why 
we must use different ways to separately generate charged-lepton- and neutrino 
masses. Besides the way done via \eqref{YD}, another way of generating lepton 
masses could be done via an effective coupling of two quartets as follows 
\begin{equation}
-{\cal L^\prime}_{f\Delta}=
\frac{Y_{\alpha\beta}^{\Delta^\prime}}{\Lambda}\bar{f}^\alpha_L({f^\beta_L})^c {\Delta^\prime},
\label{YSHeff}
\end{equation}
where $\Delta^\prime$ is a decuplet component in the decomposition of 
a tensor product 
of two quartets $\rho \otimes \chi$ or $\eta \otimes \xi$ according to the rule 
${\mathbf{4}}\otimes {\mathbf{4}}= \mathbf{6} \oplus \mathbf{10}$. Depending 
on which masses (of charged leptons or neutrinos) to be generated $\rho \otimes \chi$ or $\eta \otimes \xi$ will be chosen for ${\mathbf{4}}\otimes {\mathbf{4}}$. To express these cases we formally write $\Delta_{(1)}^\prime\sim \rho \otimes \chi$ or $\Delta_{(2)}^\prime\sim \eta \otimes \xi$. Here again the sextet component $\mathbf{6}$ in ${\mathbf{4}}\otimes {\mathbf{4}}$ is neglected as it cannot contribute to a lepton mass term. Let us denote a VEV of $\Delta^\prime$, which is either $\Delta_{(1)}^\prime$ or $\Delta_{(2)}^\prime$, as follows
\begin{equation}
\langle {\Delta^\prime}\rangle = \left[ \begin{array}{cccc}
\delta_1^\prime ~&~ 0  ~&~ \delta_2^\prime ~&~ 0\\[.9mm]
0 ~&~ 0 ~&~ 0 ~&~ \delta_4^\prime\\[.9mm]
\delta_2^\prime ~&~ 0 ~&~ \delta_3^\prime ~&~ 0\\[.9mm]
0 ~&~ \delta_4^\prime ~&~ 0 ~&~ 0
\end{array} \right], 
\label{vev10eff}
\end{equation}
where 
\begin{align}
\delta_1^\prime=\delta_2^\prime=\delta_3^\prime=0, ~\delta_4^\prime=\sigma_2w_4,
\end{align}
for $\Delta^\prime\sim \rho \otimes \chi$, or 
\begin{align}
\delta_1^\prime =u_1v_1, ~\delta_2^\prime=\dfrac{u_1v_2+u_2v_1}{2}, \delta_3^\prime=u_3v_3, ~\delta_4^\prime=0, 
\end{align}
for $\Delta^\prime\sim \eta \otimes \xi$.
\\ 

Following the latest discussions the lepton masses can be 
generated by several ways. Let us count two of them. One of the ways is 
the neutrino masses are generated by either $\Delta$ or $\Delta_{(2)}^\prime$, 
then the charged-lepton masses should be generated by an alternative decuplet. 
Another way is if $\Delta$ is involved in the generation of both the neutrino masses and charged-lepton masses, one or all of the decuplets $\Delta_{(1)}^\prime$ and $\Delta_{(2)}^\prime$ could be required to additionally contribute to the generation of either of these masses to make their total generations different from each other as required above. Here, for one of several possibilities, we will explore 
the neutrino masses generated by the decuplet $\Delta$ (with $\delta_4=0$) 
via \eqref{YD} and the charged lepton masses generated 
by $\Delta_{(1)}^\prime$  
via \eqref{YSHeff}. 
The general procedure with exchanged roles between a $\Delta$ and 
an appropriate $\Delta^\prime$ is similar and can be investigated separately with 
a feature that the VEV of $\Delta^\prime$ is adjusted by the VEV's 
of quarterts. Since the charged-lepton mass term \eqref{YSHeff} is 
independent from the neutrino one \eqref{YD} we can set at the 
beginning the charged-lepton mass matrix diagonal, i.e., 
$Y_{\alpha\beta}^{\Delta^\prime}
\sim Y_\alpha^{\Delta^\prime}\delta_{\alpha\beta}$ 
(here $\Delta^\prime\equiv \Delta_{(1)}^\prime$). 
\\

In the neutrino subspace, the coupling \eqref{YD} after $\Delta$ acquiring 
a VEV reads
\begin{align}
-{\cal L}_{M_\nu}=  
{\cal Y}_{\alpha\beta}^\Delta \left[\bar{\nu_L}^\alpha, (\bar{N}_R^\alpha)^c \right]
\left[ \begin{array}{cccc}\delta_1 & \delta_2 \\[1.5mm]
\delta_2 & \delta_3
\end{array} \right]\left[\begin{array}{c} (\nu_L^\beta)^c\\[2mm]
N_R^\beta\\
\end{array}\right],
\label{YMnu}
\end{align}
\begin{equation}
\left[\bar{\nu_L}, (\bar{N}_R)^c \right]^\alpha
{\cal M}_{\alpha\beta}\left[\begin{array}{c} (\nu_L)^c\\[2mm]
N_R\\
\end{array}\right]^\beta, 
\label{nu-mass-matrix}
\end{equation}
where the mass matrix $\cal M$ has the form 
\begin{equation}
{\cal M}=
\left[ \begin{array}{cccc}{\bf m_T} & {\bf m_D} \\[1.5mm]
{\bf m_D} & {\bf m_S}
\end{array} \right],
\label{M-matrix}
\end{equation}
in which ${\bf m_T}$, ${\bf m_D}$ and ${\bf m_S}$ are $3\times 3$ 
matrices with elements
\begin{equation}
({\bf m_T})_{\alpha\beta}=Y_{\alpha\beta}^\Delta \delta_1, 
~({\bf m_D})_{\alpha\beta}
=Y_{\alpha\beta}^\Delta \delta_2, 
~ ({\bf m_S})_{\alpha\beta}=Y_{\alpha\beta}^\Delta \delta_3,
\label{Yukawa}
\end{equation}
The magnitudes 
of the masses and their ratio depend on not only the Yukawa 
couplings but also the symmetry breaking's hierarchy to be 
discussed below. \\

\begin{figure}
	\minipage{3.95cm}
	\centering
	\includegraphics[width=4.5cm]{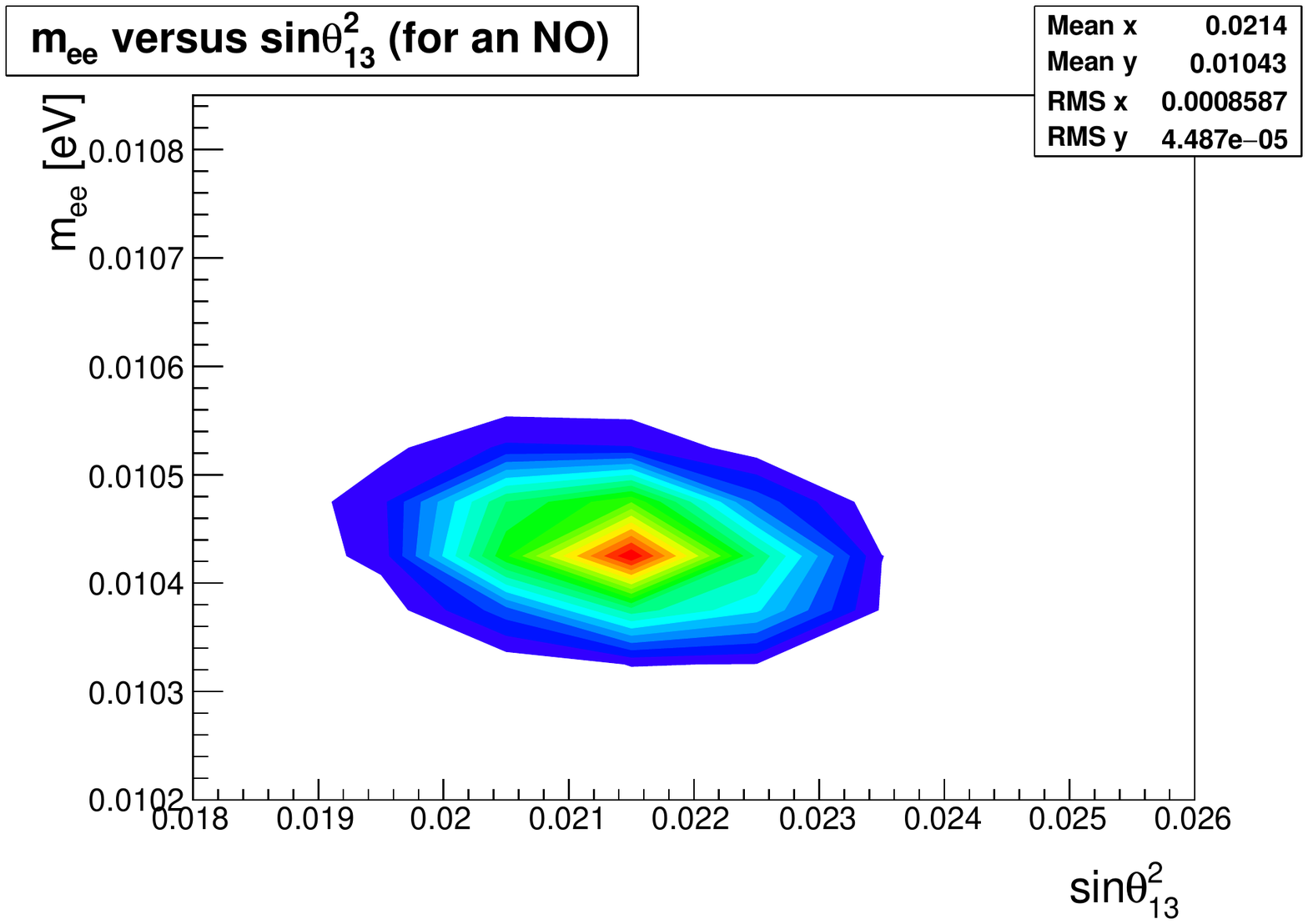} 
	\endminipage
	\hfill
	\quad
	\minipage{3.95cm}
	\includegraphics[width=4.5cm]{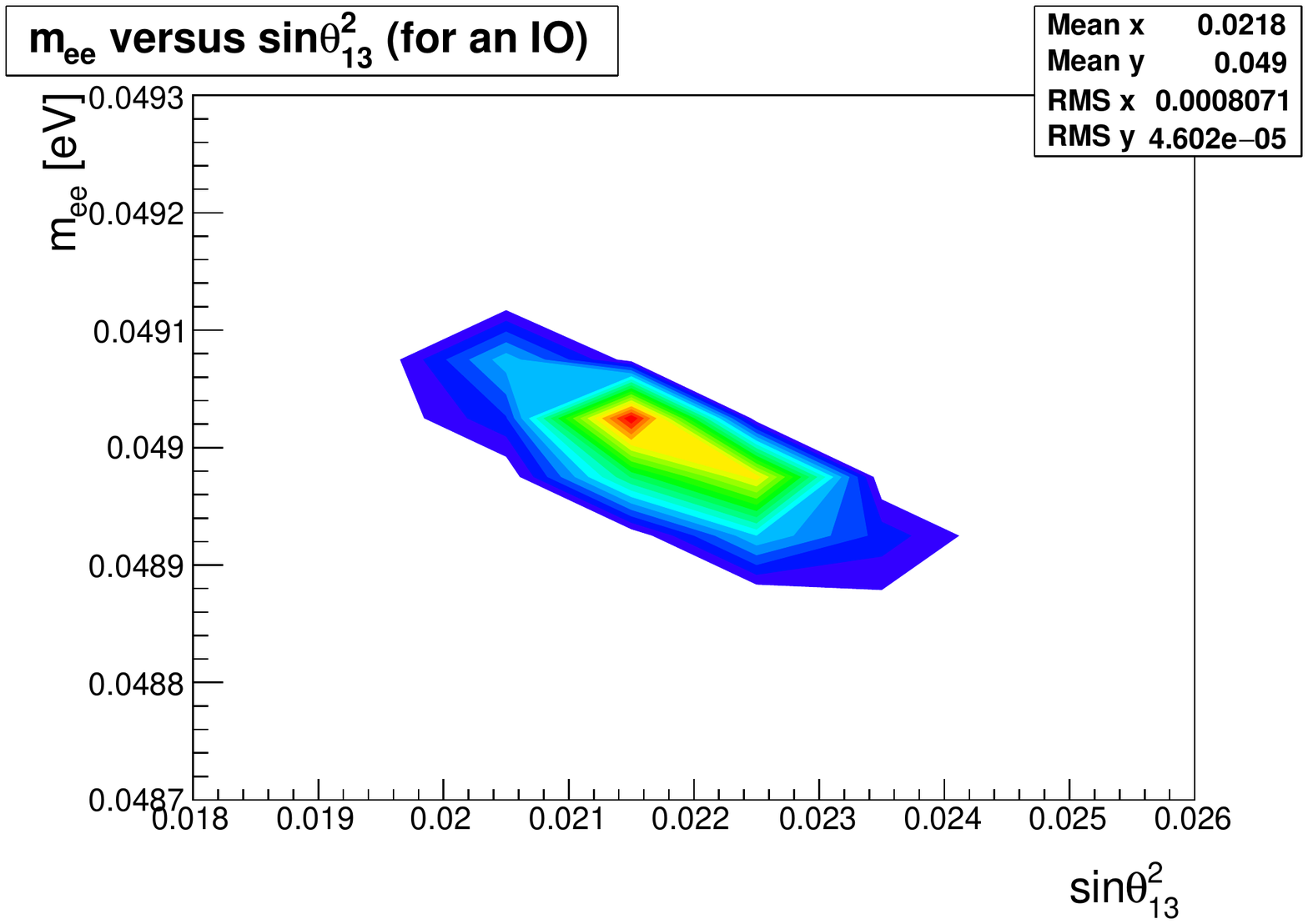}
	\endminipage
	\hfill
	\minipage{3.95cm}
	\centering
	\includegraphics[width=4.5cm]{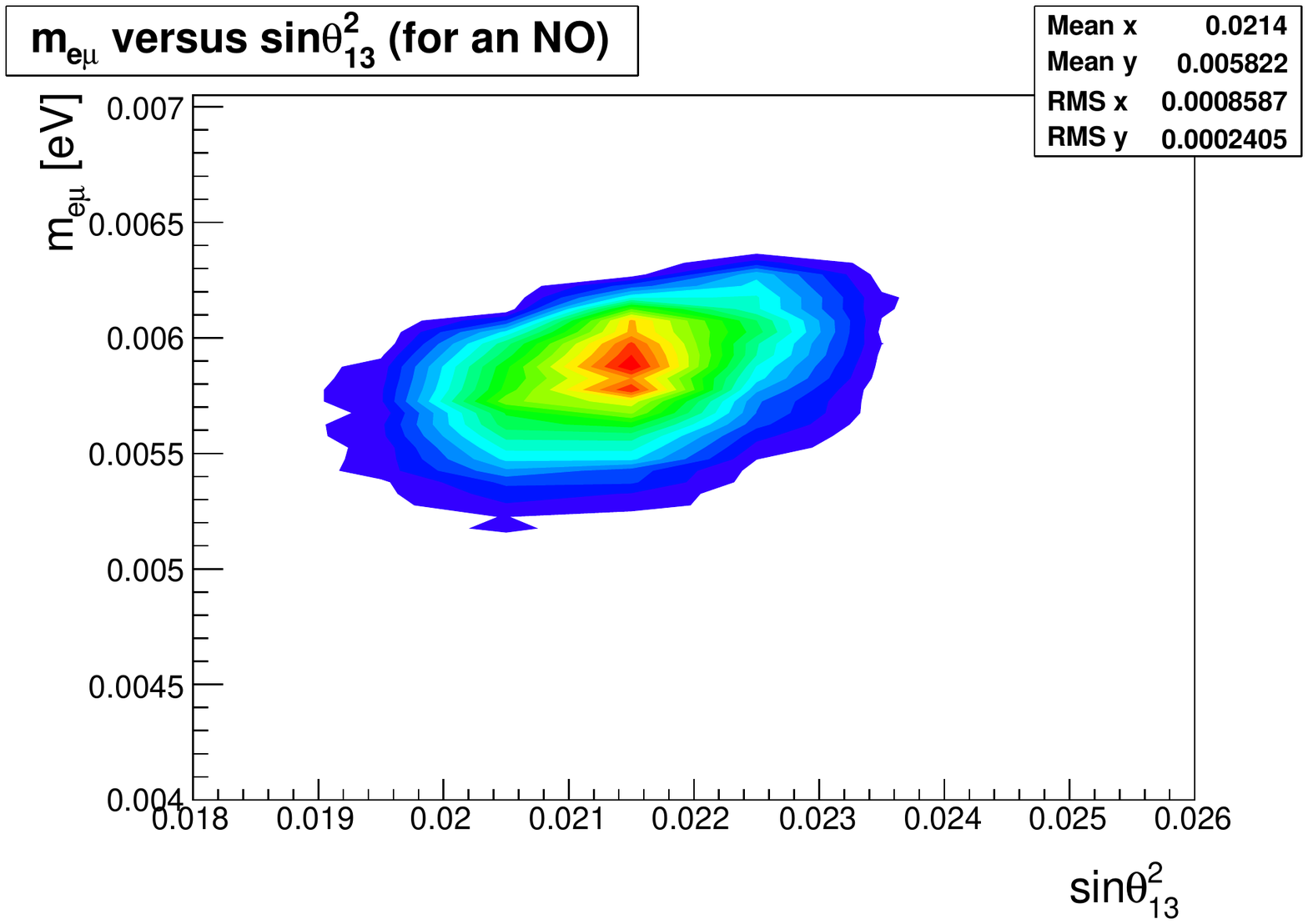} 
	\endminipage
	\hfill
	\quad
	\minipage{3.95cm}
	\includegraphics[width=4.5cm]{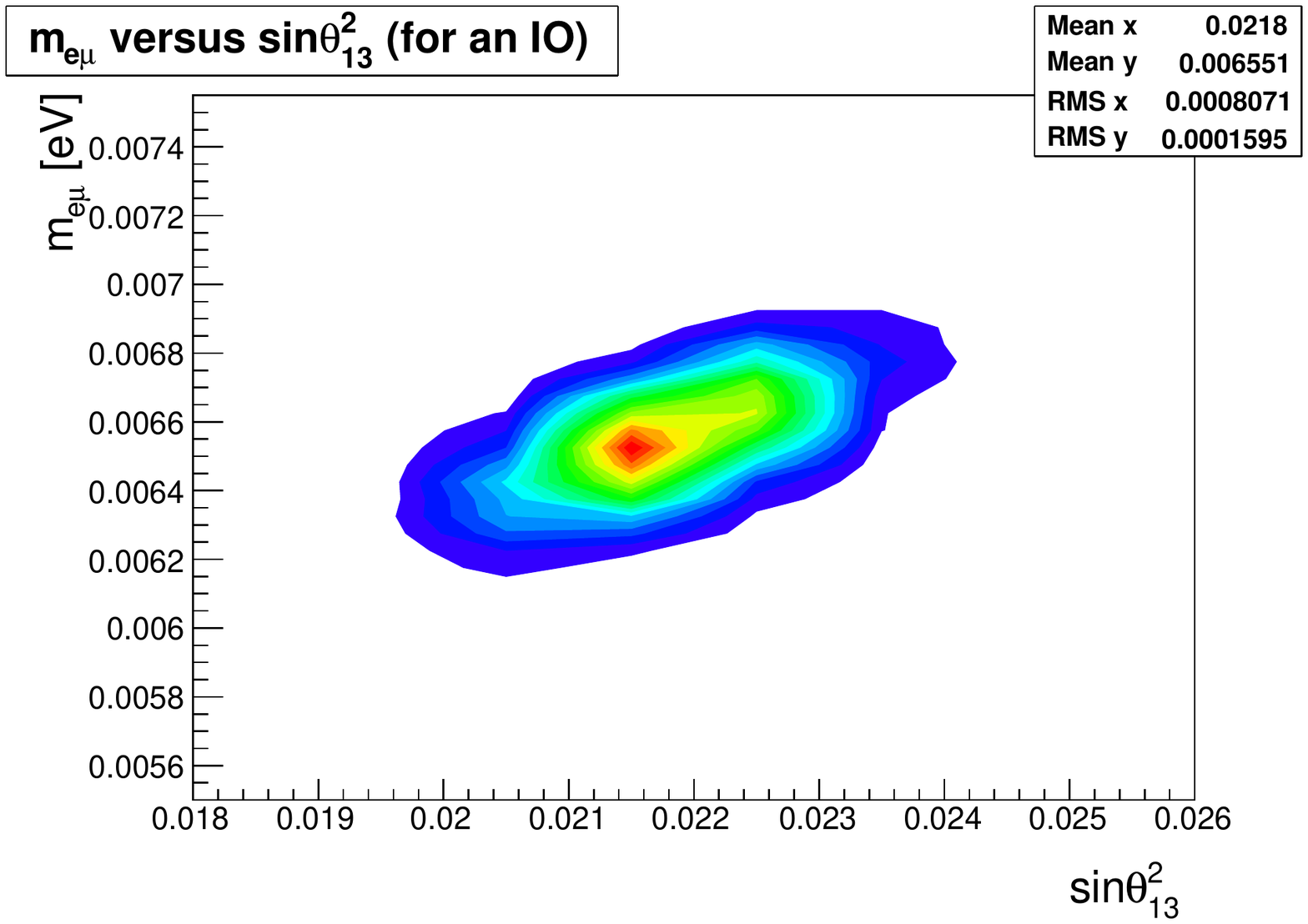}
	\endminipage
	\hfill
	\minipage{3.95cm}
	\centering
	\includegraphics[width=4.5cm]{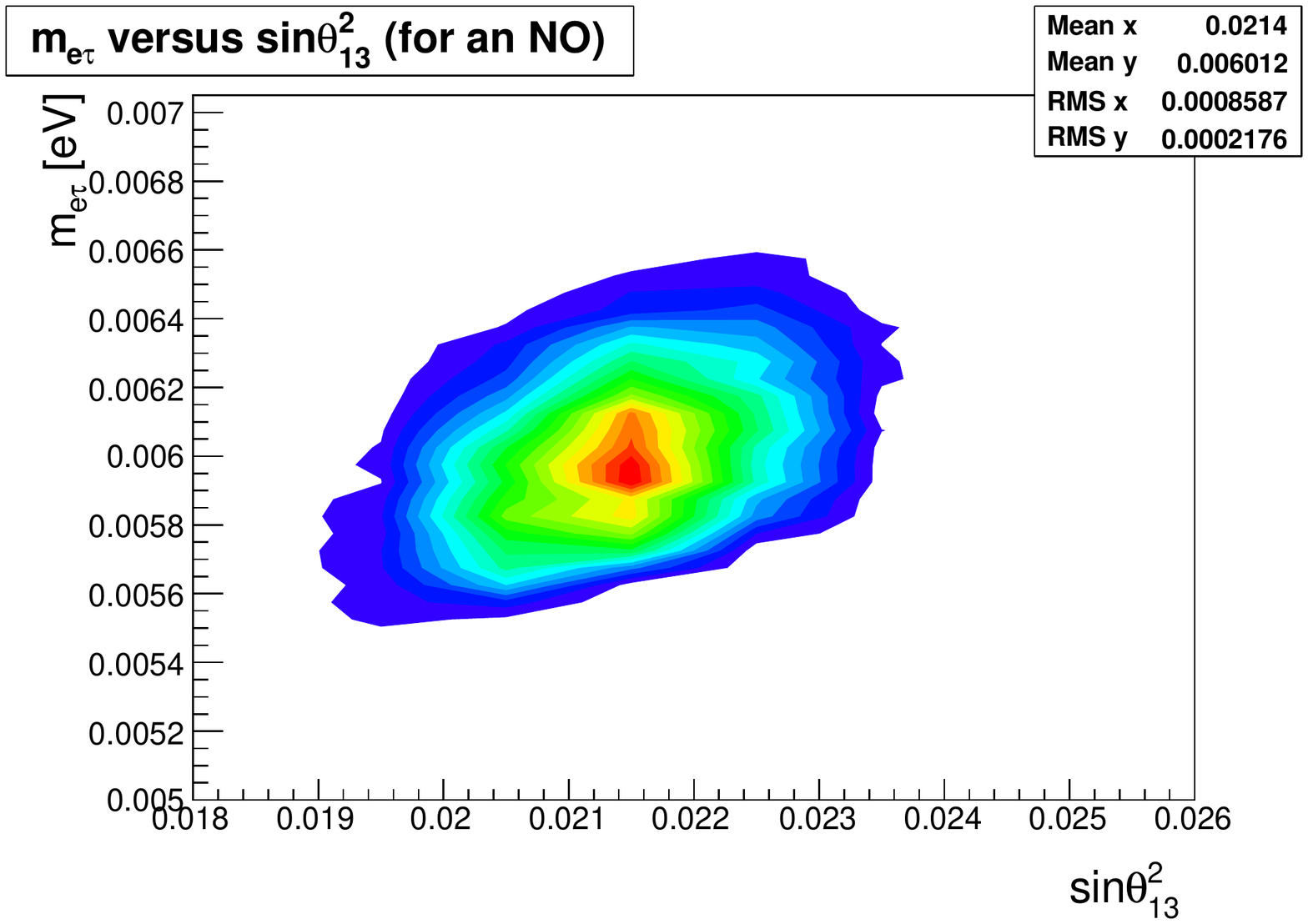} 
	\endminipage
	\hfill
	\quad
	\minipage{3.95cm}
	\includegraphics[width=4.5cm]{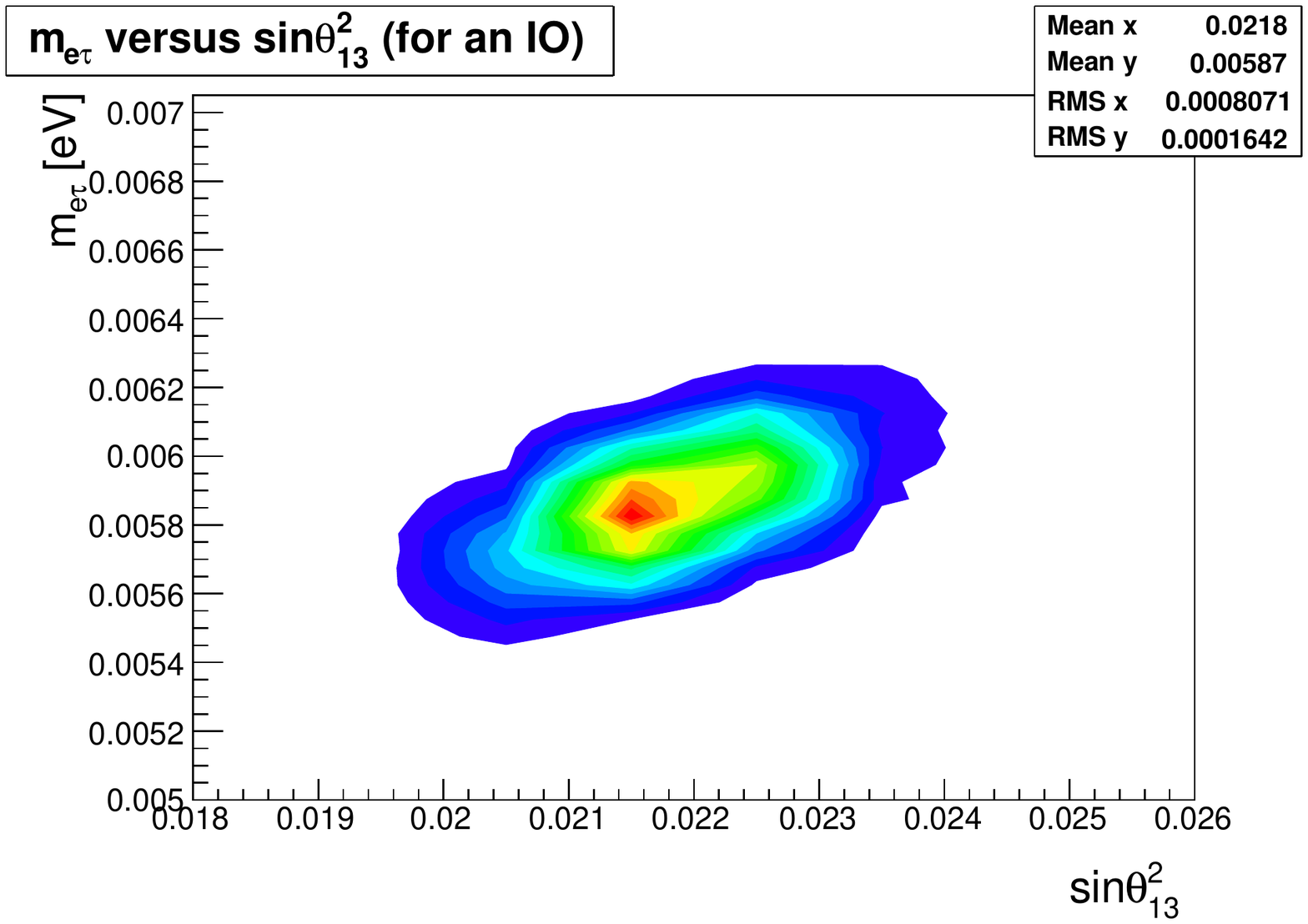}
	\endminipage
	\hfill
	\minipage{3.95cm}
	\centering
	\includegraphics[width=4.5cm]{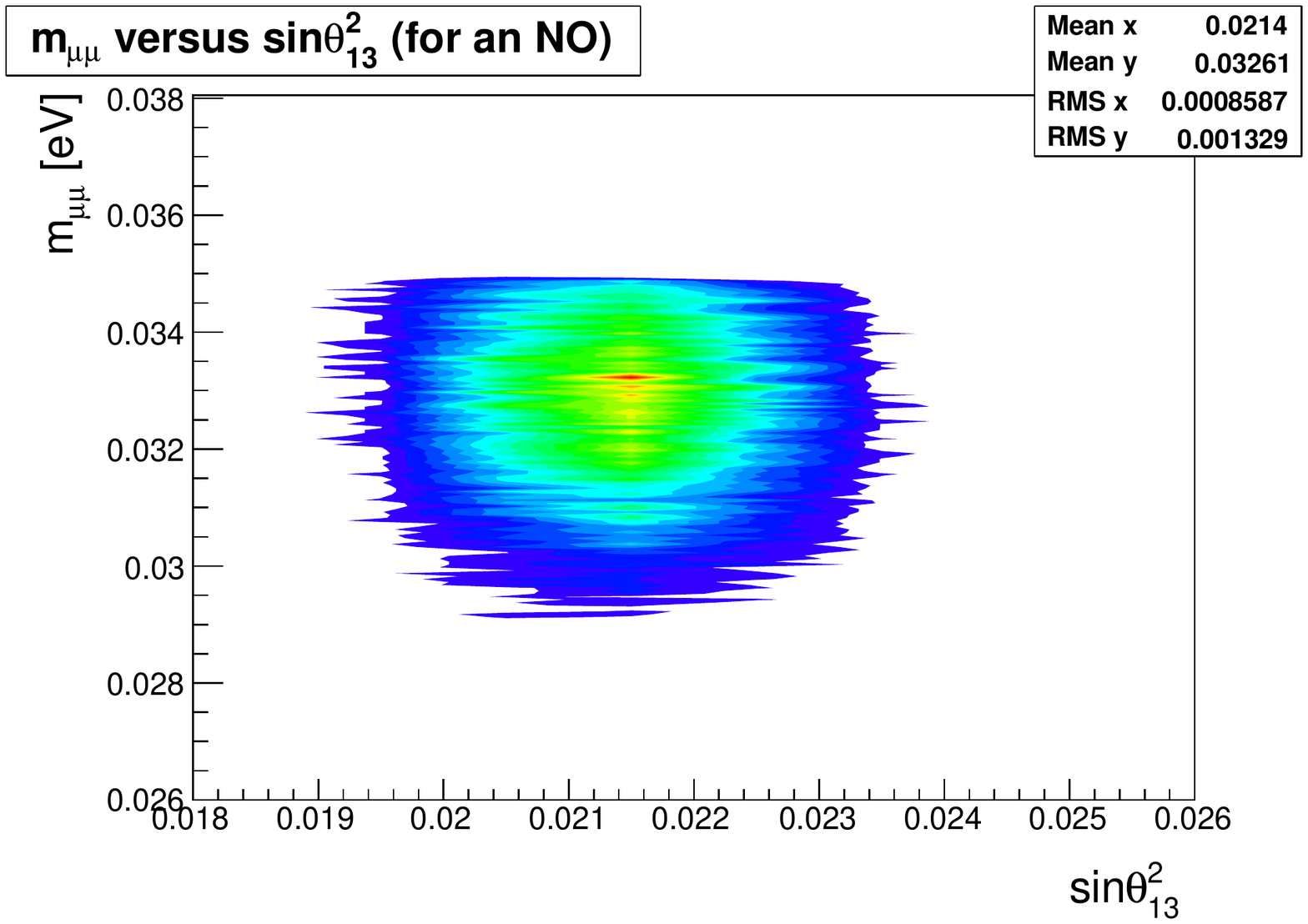} 
	\endminipage
	\hfill
	\quad
	\minipage{3.95cm}
	\includegraphics[width=4.5cm]{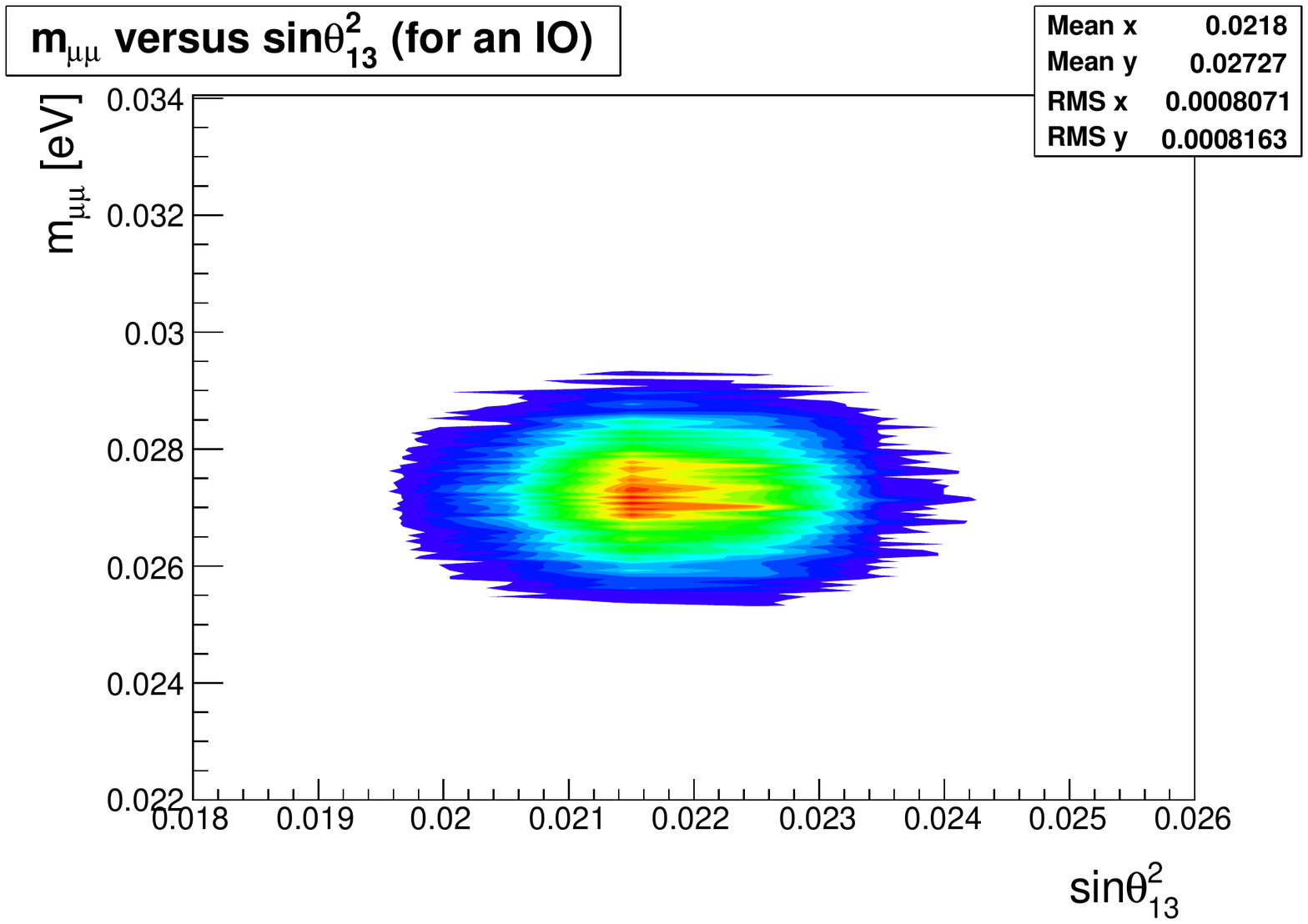}
	\endminipage
	\hfill
	\minipage{3.95cm}
	\centering
	\includegraphics[width=4.5cm]{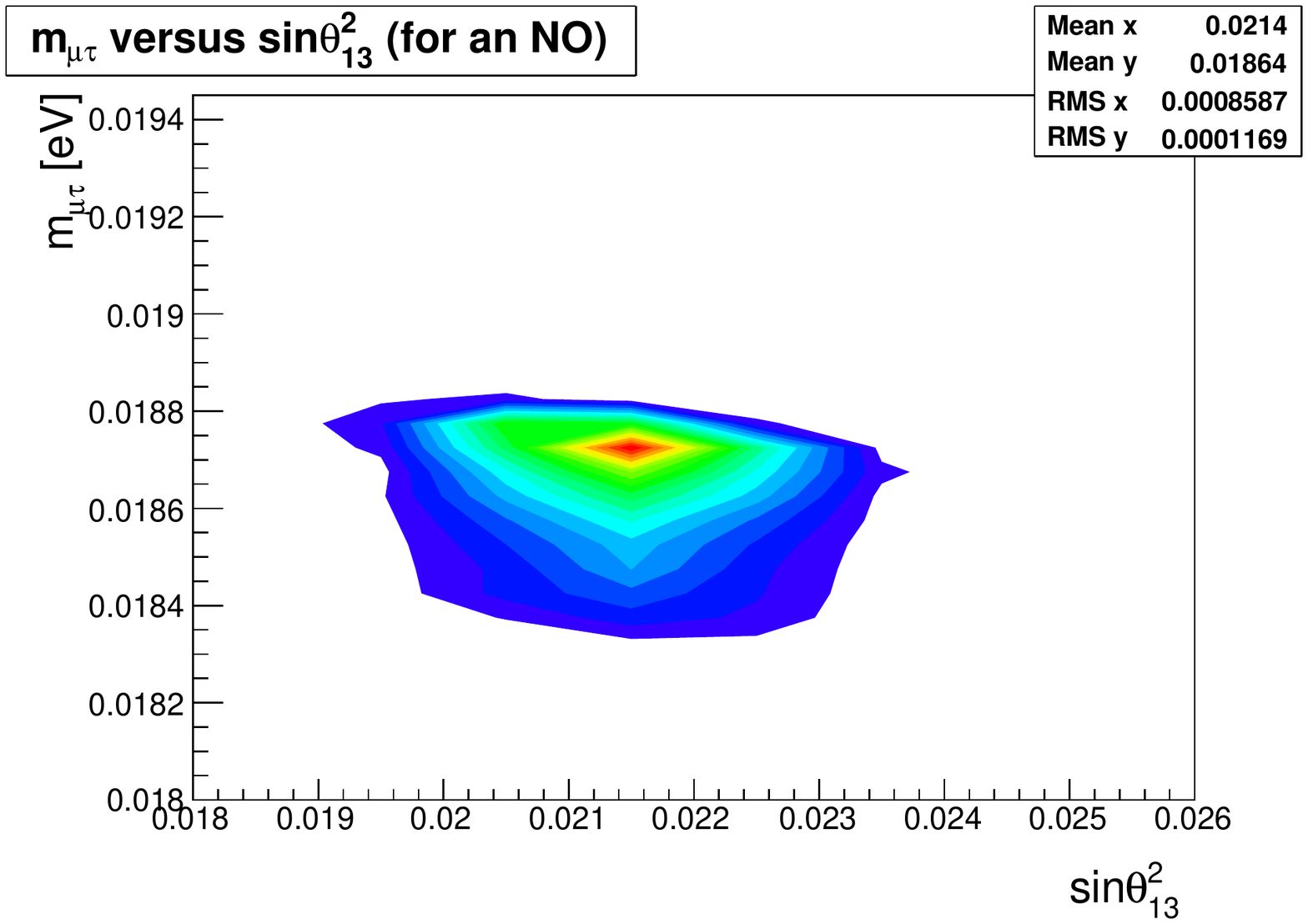} 
	\endminipage
	\hfill
	\quad
	\minipage{3.95cm}
	\includegraphics[width=4.5cm]{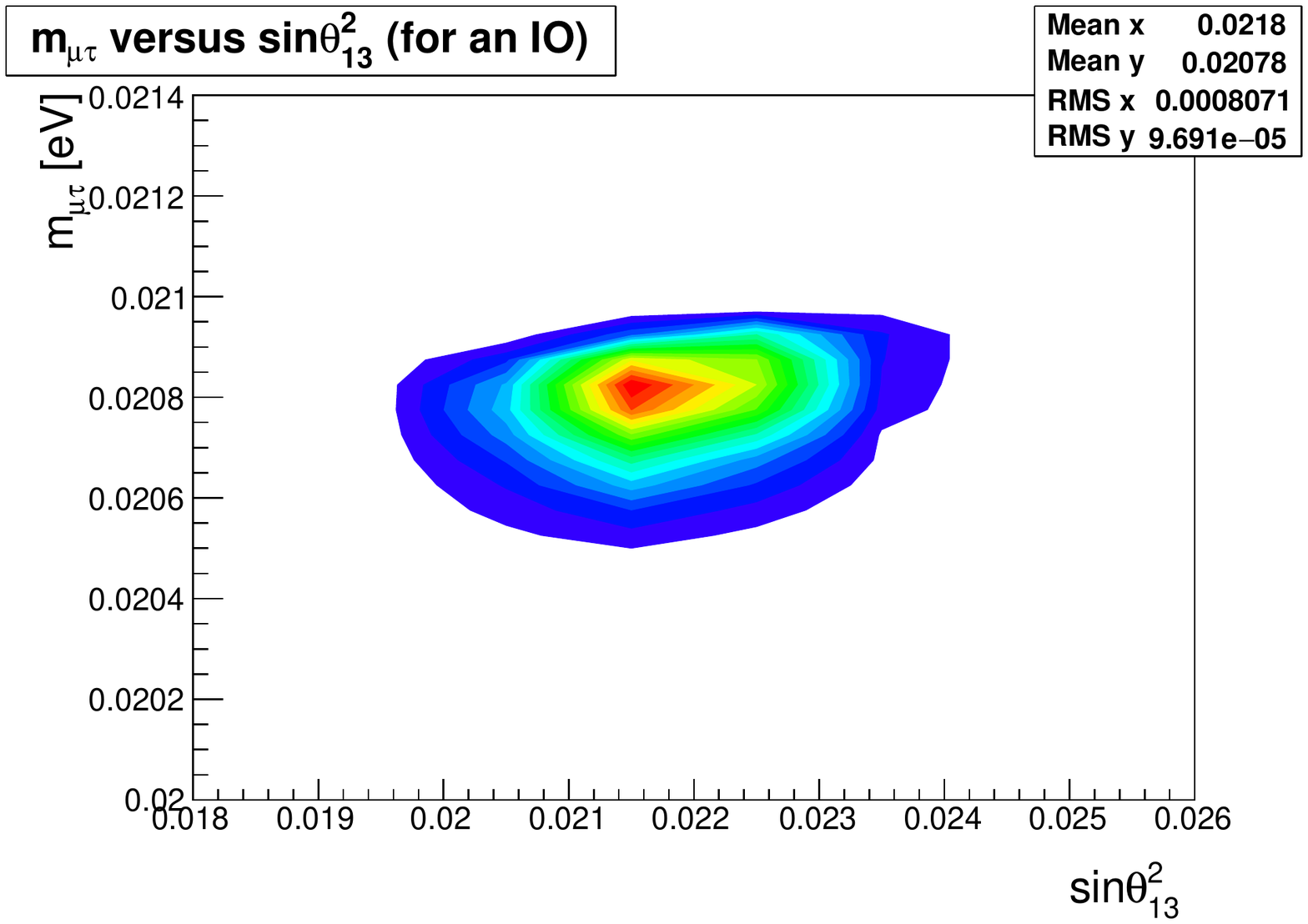}
	\endminipage
	\hfill
	\minipage{3.95cm}
	\centering
	\includegraphics[width=4.5cm]{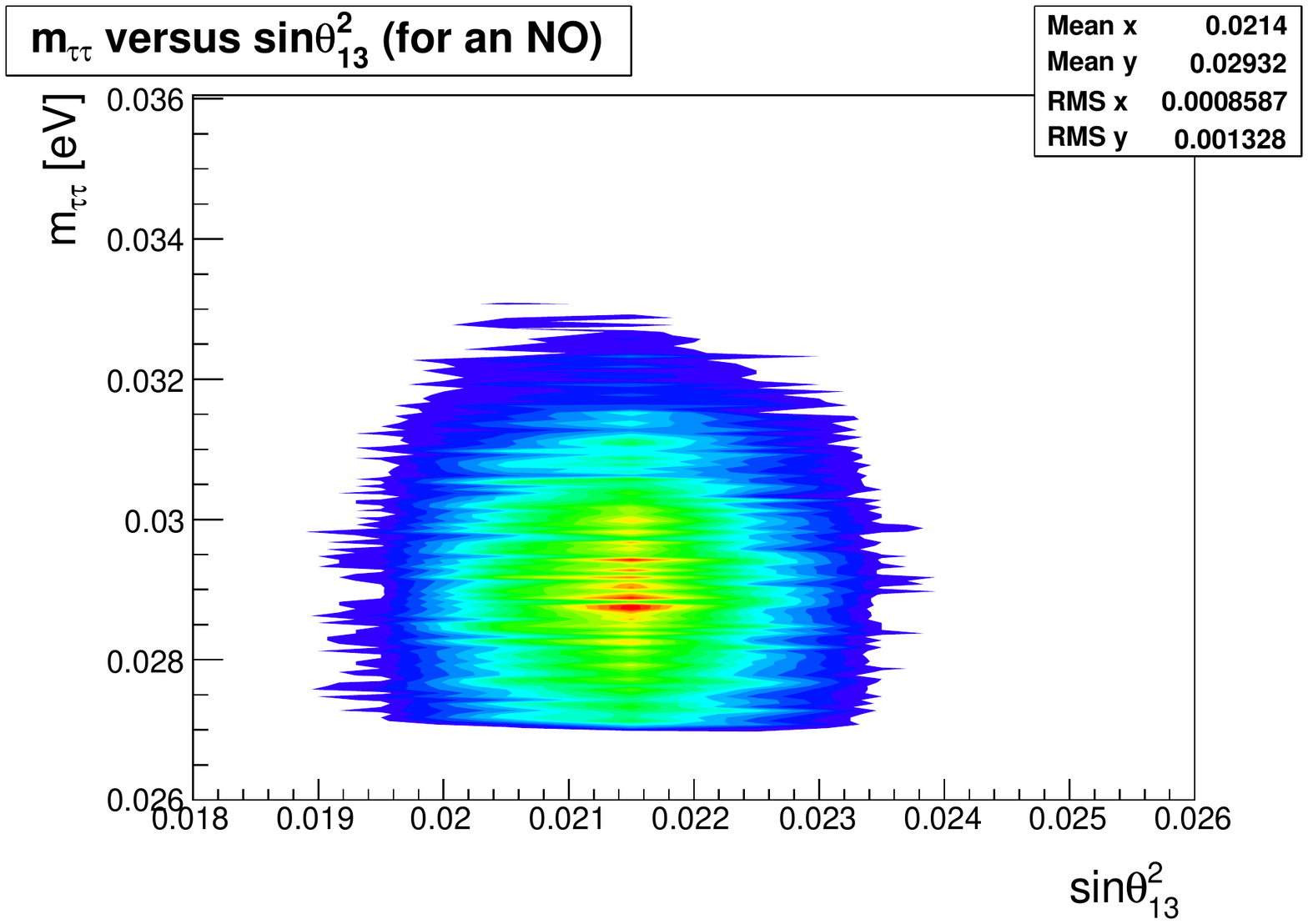} 
	\endminipage
	\hfill
	\quad
	\minipage{3.95cm}
	\includegraphics[width=4.5cm]{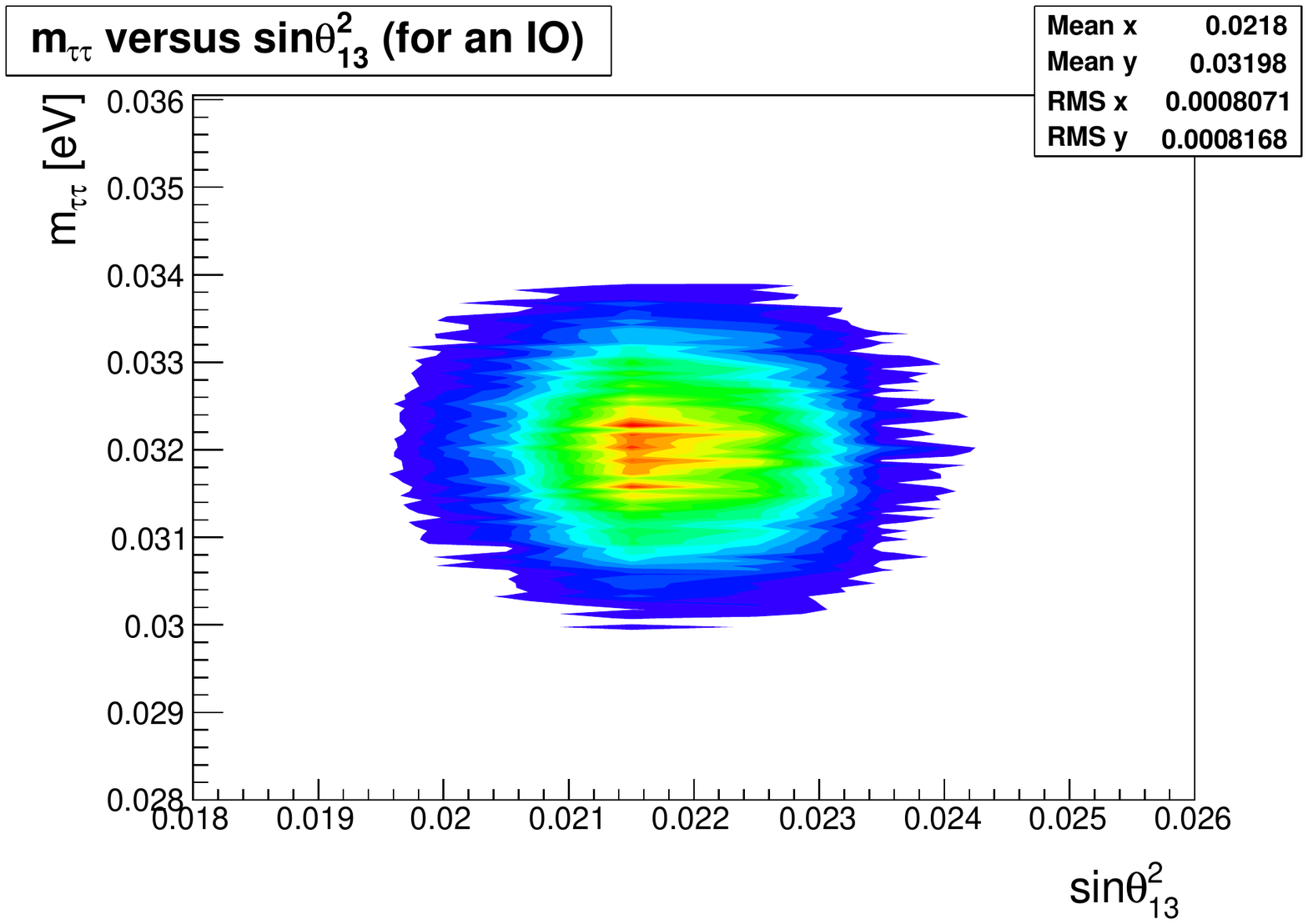}
	\endminipage
	\caption{{\label{fig1}} Distributions of $m_{\alpha\beta}\equiv (m_\nu)_{\alpha\beta}$ 
	versus $sin^2\theta_{13}$ for an NO (on the left) and for an IO 
	(on the right).}
	\hfill
	\label{mijvs13}
\end{figure}

\begin{figure}
\minipage{3.95cm}
	\centering
	\includegraphics[width=4.5cm]{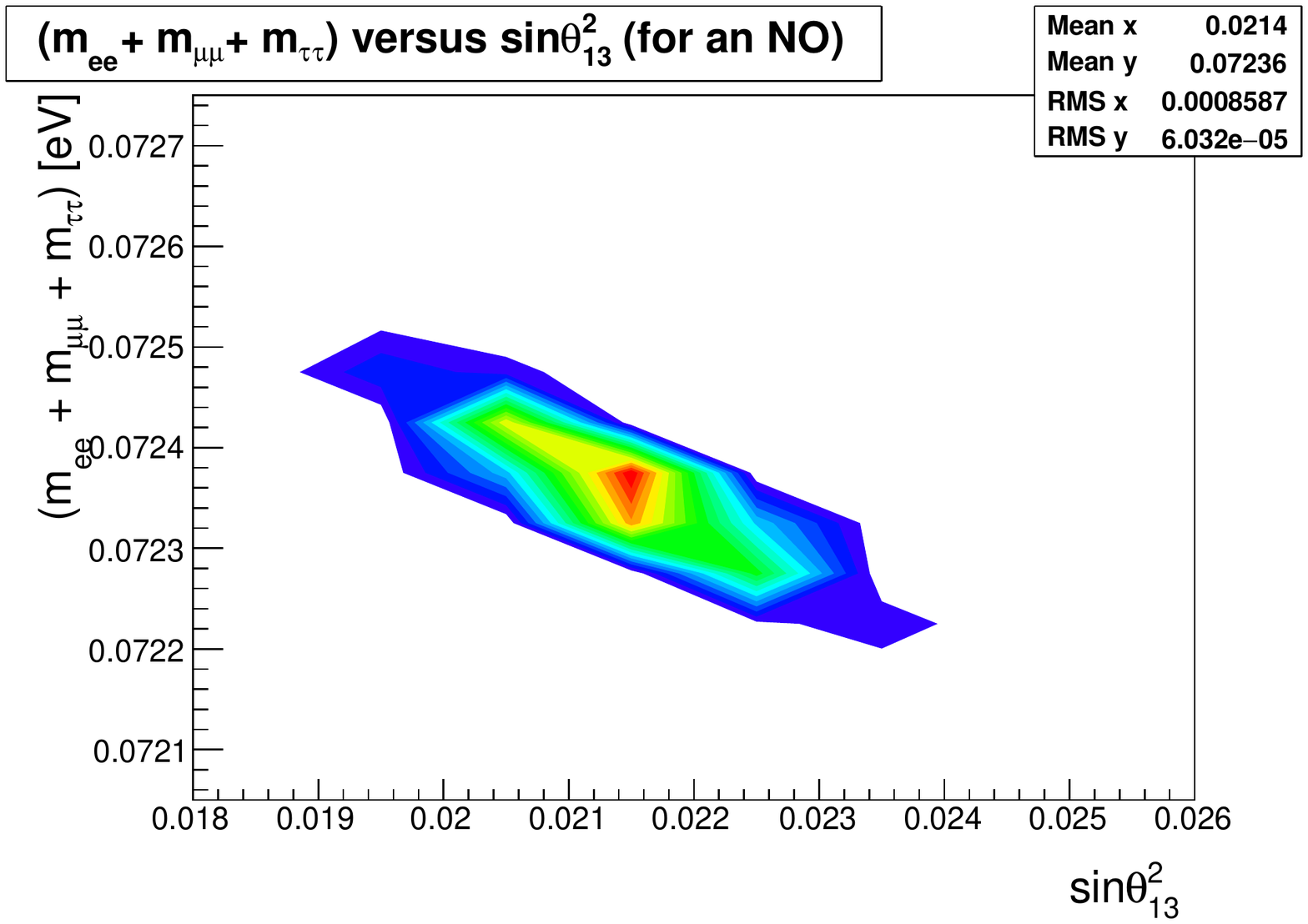} 
	\endminipage
	\hfill
	\quad
	\minipage{3.95cm}
	\includegraphics[width=4.5cm]{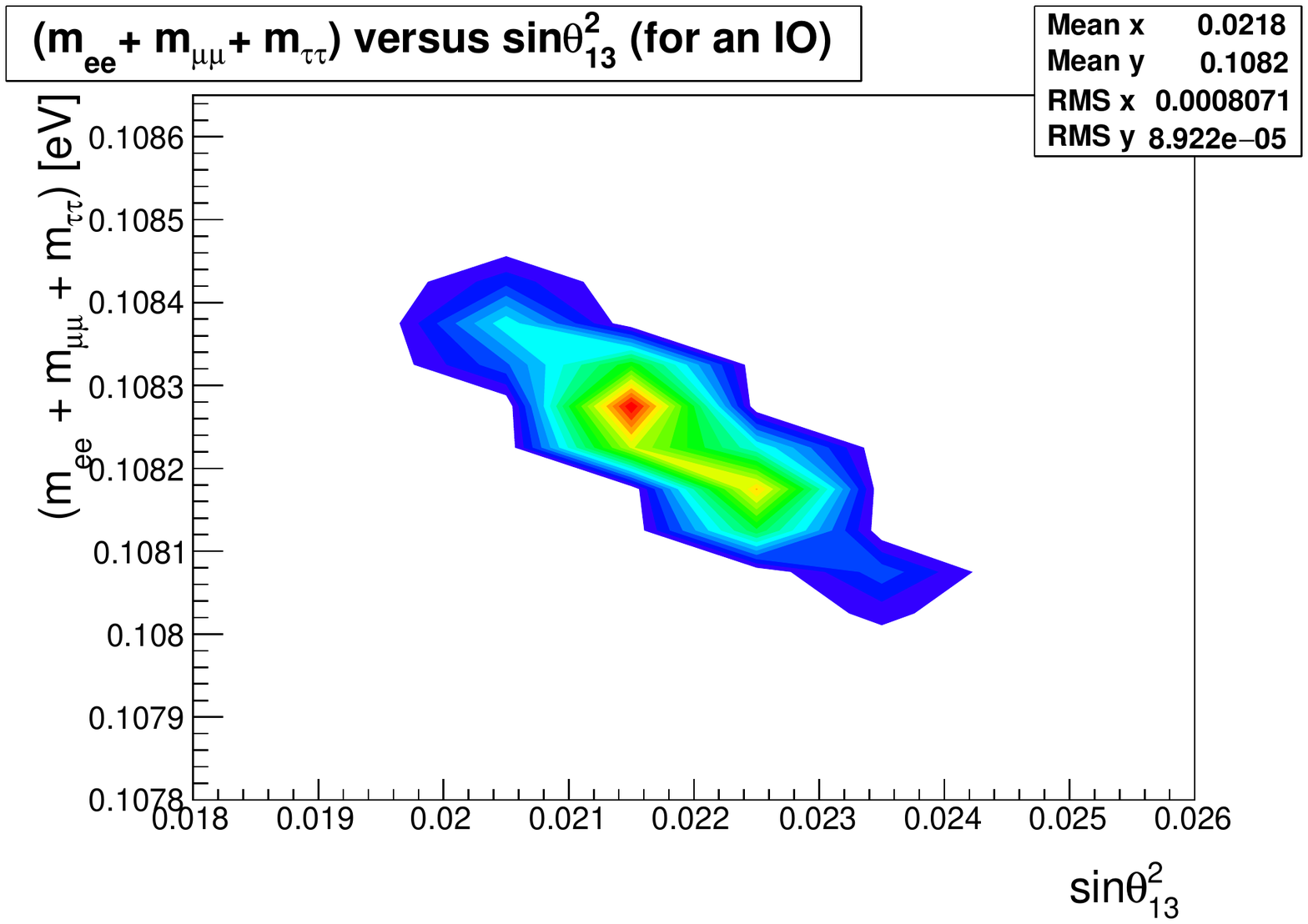}
	\endminipage
	\caption{{\label{fig2}} Distribution of $\sum m_\nu$ 
	versus $sin^2\theta_{13}$ for an NO (on the left) and for an IO 
	(on the right).}
	\hfill
	\label{miis13}
\end{figure}

The symmetry breaking scheme could be as follows: $\delta_3$ can break 
$SU(3)_L$-symmetry, thus, $SU(4)$-symmetry but not $SU(2)_L$-symmetry, 
while $\delta_2$ can break $SU(2)_L$-symmetry and $\delta_1$ (not necessary 
to be big) can be very small or zero (to break weakly or not to break a 
$U(1)$-symmetry). Thus, we should have
\begin{equation}
\delta_3\gg \delta_2\gg \delta_1.
\label{seesawlimit}
\end{equation}
That means the see-saw mechanism works for \eqref{nu-mass-matrix} leading to 
the following two eigen matrices
\begin{equation}
{\bf m}={\bf m_T}-({\bf m_D})^T({\bf m_S})^{-1}{\bf m_D}, ~~~~ {\bf M}={\bf m_S},
\label{Mseesaw}
\end{equation}
after converting the mass matrix \eqref{M-matrix} to a quasi-diagonalized form
\begin{equation}
{\cal M}_{\text{diag}}=
\left[ \begin{array}{cccc}{\bf m} & {\bf 0} \\[1.5mm]
{\bf 0} & {\bf M}
\end{array} \right].
\label{M-diago}
\end{equation}
Following \eqref{Yukawa} we get the matrix elements, denoted by $m_{\alpha\beta}$, 
of the mass matrix ${\bf m}$ of the light neutrinos 
\begin{equation}
m_{\alpha\beta}=\left[\delta_1 - {(\delta_2)^2\over \delta_3}\right]
Y_{\alpha\beta}^\Delta.
\label{lightM}
\end{equation}
The eigenvalues $m_k$, $k=1,2,3$, of the matrix ${\bf m}$ are the masses of three 
light active neutrinos (see more below). So far none of the masses $m_k$ but only 
the upper bound of their sum is known, $\sum_k m_k \equiv \sum m_\nu \leq 0.12$ eV 
\cite{Palanque-Delabrouille:2015pga,Loureiro:2018pdz,Couchot:2017pvz,Mertens:2016ihw}. 
From this bound and the current data on squared mass differences \cite{Capozzi:2018ubv} 
one can derive the bounds $0\leq  m_1/\text{eV} \leq 0.03$  for a normal neutrino mass 
ordering (NO) and $0\leq m_3/\text{eV} \leq 0.016$  for an inverse neutrino mass 
ordering (IO).
That means $m_i$ could be of the order of $10^{-1}$  eV $-$ $10^{-2}$ eV 
or lower and according to \eqref{lightM} they are related to the strength of the Yukawa 
couplings. In general the Yukawa couplings are free parameters of the model (to be 
determined experimentally directly or indirectly) but it is seen from 
Eq. \eqref{Yukawa} that they are proportional to $m_{\alpha\beta}$ which can be calculated 
numerically using the current experimental data given 
\cite{Capozzi:2018ubv}. 
Figure \ref{mijvs13} shows two-dimensional plots of distributions of $m_{\alpha\beta}$ versus 
$sin^2\theta_{13}$ for both an NO and an IO of the neutrino masses with $m_1=0.01$ eV 
(for an NO) and $m_3=0.01$ eV (for an IO), respectively, chosen as testing masses 
(other values can be chosen, but, at any case, $m_1$ and $m_3$ must be in the ranges 
$0 \leq  m_1/\text{eV} \leq 0.030 $ for an NO and 
$0 \leq m_3/\text{eV}\leq 0.016$ for an IO, as derived above), 
while other masses in each mass ordering are constrained by the squared mass differences 
\cite{Capozzi:2018ubv}. Here ten thousand events 
are created and each $m_{\alpha\beta}$ is calculated event by event as a function of mixing 
angles which are random values generated on the base of a Gaussian distribution 
having the mean (best fit) value and sigmas given in Ref. \cite{Capozzi:2018ubv}. 
From Fig. \ref{mijvs13} one can imagine how the Yukawa couplings distribute around 
their mean values (upto a scale depending on $\delta_1, \delta_2$ and $\delta_3$ as 
shown in \eqref{lightM}). Since the mean values of $m_{\alpha\beta}$ are of the 
order of $10^{-2}$ eV or smaller ($m_{\alpha\beta}\leq 10^{-2}$ eV) and $\delta_2\approx 10^2$ 
GeV (the electroweak symmetry breaking scale) the range of the Yukawa couplings could 
be $Y_{\alpha\beta}^\Delta\leq 10^{-24}\delta_3/eV$  
at the seesaw limit $\delta_1\approx 0$. They 
are stronger (weaker) for a higher (lower) $\delta_3$, therefore, an 
$SU(4)\times U(1)$-electroweak phenomenology is sensitive only if $\delta_3$ 
is high enough. The distribution shape of $\Tr(Y_{\alpha\beta}^\Delta)$ can be visualized via 
Fig. \ref{miis13} showing the distributions of $\sum m_\nu$ around the mean values 
$\overline{\sum} m_\nu\approx 0.07$ eV (for an NO) and 
$\overline{\sum} m_\nu \approx 0.11$ eV (for an IO). 
It is seen that these values are still lying within the currently established upper bound $\sum m_\nu \leq 0.12$ eV. 
%
%
If $m_1> 0.03$ eV (for an NO) or $m_3> 0.016$ eV (for an IO) the sum $\sum m_\nu$ would exceed 0.12 eV.
\\

Diagonalizing $Y_{\alpha\beta}$ we get the matrices ${\bf m_T}$, ${\bf m_D}$ 
and ${\bf m_S}$ diagonalized with eigenvalues
\begin{equation}
({m_T})_k=Y_k^\Delta \delta_1, ~({m_D})_k=Y_k^\Delta \delta_2, ~ ({m_S})_k=Y_k^\Delta \delta_3,
\end{equation}
where, $k=1,2,3$, numbering the mass eigenvalues and the mass eigenstates. 
Taking \eqref{Mseesaw}
into account this leads to the neutrino mass-states
\begin{align}
(n_{L})_k = ~ & (\nu_L)_k-{(m_D)_k\over (m_S)_k}(N_R^{~c})_k,
\nonumber\\
(N_{L})_k  = ~ & {(m_D)_k\over (m_S)_k}(\nu_L)_k+(N_R^{~c})_k,
\label{masstate}
\end{align}
corresponding respectively to the masses
\begin{align}
m_k= ~ & (m_T)_k-{(m_D)_k^2\over (m_S)_k} \equiv 
\left[\delta_1-{(\delta_2)^2\over \delta_3}\right]Y^\Delta_k,
\nonumber\\
M_k= ~ & (m_S)_k\equiv Y^\Delta_k \delta_3,
\label{massseesaw}
\end{align}
where the notation $N_R^{~c}\equiv (N_R)^c$ is used.
Note that at $\delta_1\approx 0$ the ratio $m_k/M_k$ becomes universal as it depends on 
neither $k$ nor the coupling coefficients $Y^\Delta_k$ but the ratio $\delta_2/\delta_3$:
\begin{equation}
{m_k\over M_k} \approx \left(m_D\over m_S\right)^2 = \left(\delta_2\over \delta_3\right)^2.
\label{massratio}
\end{equation}
That means a ratio can be predicted by knowing the other one.
Using \eqref{massratio} we can rewrite \eqref{masstate} 
in the form
\begin{align}
(n_L)_k = ~ & (\nu_L)_k-\sqrt{m_k\over M_k} ~(N_R^{~c})_k,
\nonumber\\
(N_L)_k  = ~ & \sqrt{m_k\over M_k}
~(\nu_L)_k+(N_R^{~c})_k,
\label{mix2}
\end{align}
or
\begin{align}
(n_L)_k = ~ & (\nu_L)_k-{\delta_2\over \delta_3} ~(N_R^{~c})_k,
\nonumber\\
(N_L)_k  = ~ & {\delta_2\over \delta_3}
~(\nu_L)_k+(N_R^{~c})_k.
\label{hmix}
\end{align} 
Solving the system of equations \eqref{hmix} for $(\nu_L)_k$ we get 
\begin{align}
	(\nu_L)_k = {1\over 1+(\delta_2/ \delta_3)^2}(n_L)_k +   
	{\delta_2/\delta_3\over 1+ (\delta_2/ \delta_3)^2} ~(N_L)_k,\label{Nmix}
\end{align}
or, as $\delta_2\ll \delta_3$, 
\begin{equation}
	(\nu_L)_k \approx (n_L)_k +  {\delta_2\over \delta_3}~(N_L)_k.\label{Nmix}
	\end{equation}
In the flavour basis, the neutrinos $(\nu_L)_\alpha$, $\alpha=e,\mu,\tau$, 
have the following general mixing 
\begin{align}
(\nu_L)_\alpha \approx ~&  \sum_{k=1}^3 U_{\alpha k} (n_L)_k +   
\sum_{k=1}^3 \Theta_{\alpha k} ~(N_L)_k,\label{Nmix}
\end{align}
where, $U_{\alpha k}$ is the PMNS matrix, and 
$\Theta_{\alpha k}\approx  {\delta_2\over \delta_3}U_{\alpha k}$.
That means, the flavour neutrinos in general are mixtures between light 
active neutrinos and heavy neutrinos which now are objects of increasing 
intensive search. Since mixing angles with heavy neutrinos are very small 
they are often neglected but when experiments, especially those searching 
for heavy neutrinos, become more and more sensitive and precise they should 
be taken into account.\\

At the present we do not know the exact bounds of $m_S$, which can spread from 
a relatively low scale at keV (or lower) to a very high energy scale near the 
Planck mass, but we know from the experiment and cosmological constraints 
\cite{Palanque-Delabrouille:2015pga, Loureiro:2018pdz,Couchot:2017pvz,Mertens:2016ihw,Aghanim:2018eyx} the upper bound of the active
neutrions masses $m < 10^{-1}eV$. Thus, $m/M$ can be calculated for a given $M$, 
for example, $m/M<10^{-4}$ (that means $\delta_2/\delta_3<10^{-2}$) for $M$ at 
a keV scale and $m/M<10^{-10}$ (that means $\delta_2/\delta_3<10^{-5}$) for $M$ 
at a GeV scale. The latter estimations do not contradict with the upper bounds 
of $|\Theta|^2$ established for several (current and future) experiments for 
$M$ of a few GeV's \cite{Asaka:2016rwd,Dev:2013wba}. As is well known that an 
existence of Majorana neutrinos violates the lepton number consevation rule. \\

At the $SU(2)_L$-breaking scale $\delta_2\sim 10^2$ GeV there should be keV heavy 
neutrinos ($M\sim 10^3$ eV) if the $SU(3)_L$ (or $SU(4)$) is broken at 
$\delta_3\sim 10$ TeV scale. The existence of eV or MeV heavy neutrinos requires 
the breaking scale $\delta_3 \sim 10^0$ TeV or $\delta_3 \sim 10^3$ TeV, respectively. 
The possible existence of the light heavy neutrinos (with masses, for example, 
at eV-, keV-, MeV scale) is very interesting not only in particle physical aspect 
but also in the astro-particle physical and the cosmological aspects (see, for example, \cite{Dinh:2006ia,Adhikari:2016bei} and  references therein).
\section{Conclusion}

\noindent

\begin{figure}
\begin{center}	
\includegraphics[scale=.45]{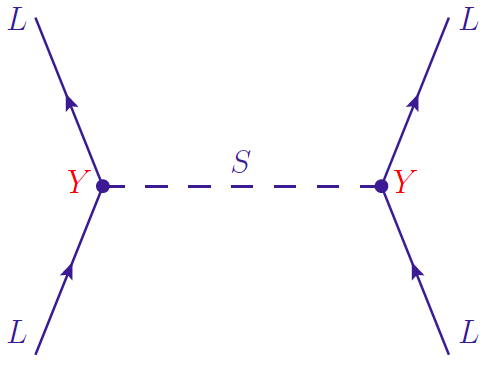} 
\caption{Scalar-exchanging lepton scattering.}
	\end{center}	
	\label{LLS}	
\end{figure}

It is well known from the experiment 
\cite{Fukuda:1998tw,Fukuda:1998ub,Fukuda:1998mi,Ahmad:2001an,Ahmad:2002jz,Ahmad:2002ka} 
that neutrinos have masses but they are very small, for example, some combined particle 
physics data and cosmological probes give an upper bound of the sum of neutrino 
masses as $\sum m_\nu < 0.12$ eV (95\% C.L.) \cite{Palanque-Delabrouille:2015pga, 
Loureiro:2018pdz,Couchot:2017pvz,Mertens:2016ihw,Aghanim:2018eyx}. 
Many theoretical models and mechanisms have been suggested to predict or  explain this experimental 
fact but none of them is completely satisfactory. In this paper we have suggested one more way of 
neutrino mass generation through spontaneous breaking of an extended $SU(4)\times U(1)_X$ electroweak 
symmetry by an $SU(4)$-decuplet scalar acquiring a VEV without using fundamental quartet scalars which 
cannot generate neutrino masses (and charged-lepton masses) directly. There are limits in which the 
seesaw mechanism can be realized. Depending on these limits the new (heavy) neutrinos added to the 
active (light) neutrinos may have masses at different ranges none of which has been so far excluded 
from the experiment. It should be noted that if the physics of the 3-4-1 model is at around TeV scale, such as that of the LHC and near future accelerators, there may exist light heavy neutrinos (at an eV-keV scale) attracting great interest in particle physics and cosmology (see, for example, 
\cite{Adhikari:2016bei,Bertuzzo:2018itn,Aguilar-Arevalo:2017vlf,Liventsev:2013zz, Andres:2017daw}). 
For $M_N$ at the order of $10^2$ GeV considered in  
\cite{Dev:2013wba} the scale of physics of 3-4-1 model, if existing, would be too high in order 
to be discovered at the LHC and other present accelerators. The neutrino masses depend on not 
only the symmetry breaking scales but also Yukawa couplings (to the decuplet) the distribution 
shapes of which are shown in Figs. \ref{mijvs13} and \ref{miis13} via the model-independent 
distributions of the neutrino mass matrix elements. Thus, these Yukawa couplings can be 
determined if the symmetry breaking scales are known and vice versa. 
More precisely, the Yukawa couplings $Y_{\alpha\beta}^\Delta$ and their trace $\Tr(Y_{\alpha\beta}^\Delta)$ can be determined from the distributions in Figs. \ref{mijvs13} and \ref{miis13} upto the factor $\left[\delta_1 - \dfrac{(\delta_2)^2}{\delta_3}\right]$. 
These distributions are derived by using an experimental data for squared mass differences and mixing angles (it means that the PMNS matrix is known) as well as a mass input respecting the mass upper bound $\sum m_\nu \leq 0.12$ eV
and,  
vice versa, the PMNS matrix can be determined via the model's parameters fixed by other independent ways, for example, via non-neutrino processes like thoses for S-exchanging L-L scatterings schematically depicted in Fig. \ref{LLS}
(here the symbol ``L" stands for a {\it charged lepton} and the symbol ``S" stands for a {\it scalar} from a decuplet). It is a quite long but very interesting work being currently investigated.
\\

As discussed in \cite{Ky:2005yq}, besides the seesaw limit $\delta_1\approx 0$, other limits in the 
mass term \eqref{nu-mass-matrix} can be considered: the pure Majorana limit ($m_D=0$), the Dirac 
limit ($m_T=m_S=0$), the pseudo-Dirac limit ($m_T\ll m_D$ and $m_S\ll m_D$), etc. It can be seen 
that the pure Majorana limit breaks the present structure of the neutrino mass term to a left-right 
$SU(2)_L\times SU(2)_R$ structure which can be a subject of a later investigation. We would like 
to stress that in the present paper the seesaw mechanism is applied, to our knowledge, for the first time to the 3-4-1 model with a scalar decuplet. \\

Finally, it is worth noting that here we have used a fundamental scalar decuplet for neutrino mass 
generation but using a decuplet composed of quartets, or using an efective coupling of the latter, 
is another possibility for generating lepton masses including masses of neutrinos and charged leptons. This research is in progress. 
\section*{Acknowledgement}
\noindent

This work is funded by Vietnam's National Foundation for Science and Technology 
Development (NAFOSTED) under grant N\textsuperscript{\underline{o}} 103.99-2018.45. The authors would like to thank Jean-Marie Frere for useful discussions.  


\end{document}